\documentclass{elsart5p}

\usepackage{graphics}
\usepackage{graphicx}
\usepackage{amssymb}
\usepackage[squaren,thinspace,thinqspace]{SIunits}	

\newcommand{\dd}{\mathrm{d}}					 							
\newcommand{\YRS}{YbRh$_2$Si$_2$}									
\newcommand{\YCxS}{Yb(Rh$_{1-x}$Co$_{x}$)$_2$Si$_2$}			
\newcommand{\YCdS}{Yb(Rh$_{0.97}$Co$_{0.03}$)$_2$Si$_2$}		
\newcommand{\YCsS}{Yb(Rh$_{0.93}$Co$_{0.07}$)$_2$Si$_2$}		
\newcommand{\TN}{\ensuremath{T_{\mathrm N}}}						
\newcommand{\TL}{\ensuremath{T_{\mathrm L}}}						
\newcommand{\Tstar}{\ensuremath{T^{\star}}}						
\newcommand{\mK}{\milli\kelvin}										
\newcommand{\Neel}{N{\'e}el}											

\begin{document}

\begin{frontmatter}

\title{Electrical resistivity of \YCxS\ single crystals \\
at low temperatures}

\author[a]{S. Friedemann\corauthref{S. Friedemann}}
	\ead{Sven.Friedemann@cpfs.mpg.de}
\author[a]{N. Oeschler}
\author[a]{C. Krellner}
\author[a]{C. Geibel}
\author[a]{F. Steglich}
\address[a]{Max Planck Institute for Chemical Physics of Solids, Noethnitzer Strasse 40, 01187 Dresden, Germany}

\corauth[S. Friedemann]{Corresponding author. Tel: +49(351)4646-3219 fax: -3232}

\begin{abstract}
We report low-temperature measurements of the electrical resistivity of \YCxS\ single crystals with $0\leq x \leq 0.12$. The isoelectronic substitution of Co on the Rh site leads to a decrease of the unit cell volume which stabilizes the antiferromagnetism. Consequently, the antiferromagnetic transition temperature increases upon Co substitution. For $x = 0.07$  Co content a subsequent low-temperature transition is observed in agreement with previous susceptibility measurements and results on \YRS\ under hydrostatic pressure. Above the \Neel\ transition the resistivity follows a linear temperature dependence of a non-Fermi liquid similar to that of \YRS. 

\end{abstract}


\end{frontmatter}

\YRS\ is a prototypical heavy fermion compound close to a magnetic quantum critical point \cite{Trovarelli2000b}. It exhibits an antiferromagnetic transition at the \Neel\ temperature $\TN = \unit{70}\mK$. \TN\ is continuously suppressed to zero temperature by small magnetic fields $B$ of about \unit{0.7}\tesla\ and \unit{0.06}\tesla\ applied parallel and perpendicular to the crystallographic $c$-axis, respectively. In the vicinity of this field-tuned quantum critical point pronounced non-Fermi liquid behavior has been observed. This manifests itself for instance in a linear temperature dependence of the resistivity $\rho$. For the heavy quasiparticle mass a stronger than logarithmic divergence was found \cite{Custers2003}, whereas the Gr{\"u}neisen ratio diverges as $T^{-0.7}$ \cite{Kuchler2003}. Both observations are in contrast to the predictions of the conventional spin-density wave description for the quantum critical point in this compound \cite{Millis1993,Moriya1995}. Furthermore, the Hall effect was found to show a crossover along a line $\Tstar(H)$ which extrapolates to a discontinuous jump at the quantum critical point if extrapolated to the zero-temperature limit \cite{Paschen2004}. From this finding a reconstruction of the Fermi surface at the quantum critical point was inferred. This is not expected in the conventional scenario. Moreover, it provides evidence for the ''Kondo-breakdown`` scenario, where the Fermi surface is expected to abruptly change at the quantum critical point \cite{Si2001}. The Hall crossover is accompanied by features in various other thermodynamic and transport properties, establishing an energy scale $T^*(B)$ besides $\TN(B)$ and $T_{\mathrm {LFL}}$, the Landau Fermi-liquid temperature \cite{Gegenwart2007}. \Tstar\ may then be associated with the break-up of the quasiparticles. 

An outstanding issue is to clarify the link of the various energy scales to the quantum criticality. For this purpose, a systematic investigation of the volume dependence of the temperature-field phase diagram of \YRS\ is needed. Here, the isoelectronic substitution of Rh with smaller Co is expected to change the strength of the magnetic interactions via the volume decrease. This effect was indeed observed in stoichiometric \YRS\ under hydrostatic pressure which revealed an increase of \TN\ with increasing pressure \cite{Mederle2002}. In an initial study of the magnetic susceptibility $\chi$ of \YCxS ,  an increase of \TN\ with increasing Co content was uncovered \cite{Westerkamp2008} which is consistent with the pressure data.

Here, we report low-temperature resistivity measurements on \YCxS\ single crystals. The resistivity was measured in zero magnetic field down to \unit{19}\mK\ in a $^3$He-$^4$He dilution refrigerator using standard four-point lock-in technique at low frequencies. Only a very tiny out-of-phase signal of less than 1\usk\% was observed, which proves the good quality of the spot-welded electrical contacts. Very high resolution was realized using low-temperature transformers. The measurements on \YCxS\ with $x=0.12$ were conducted in a physical properties measurement system using the $^3$He setup.

\begin{figure}
		\includegraphics[width=7.65cm]{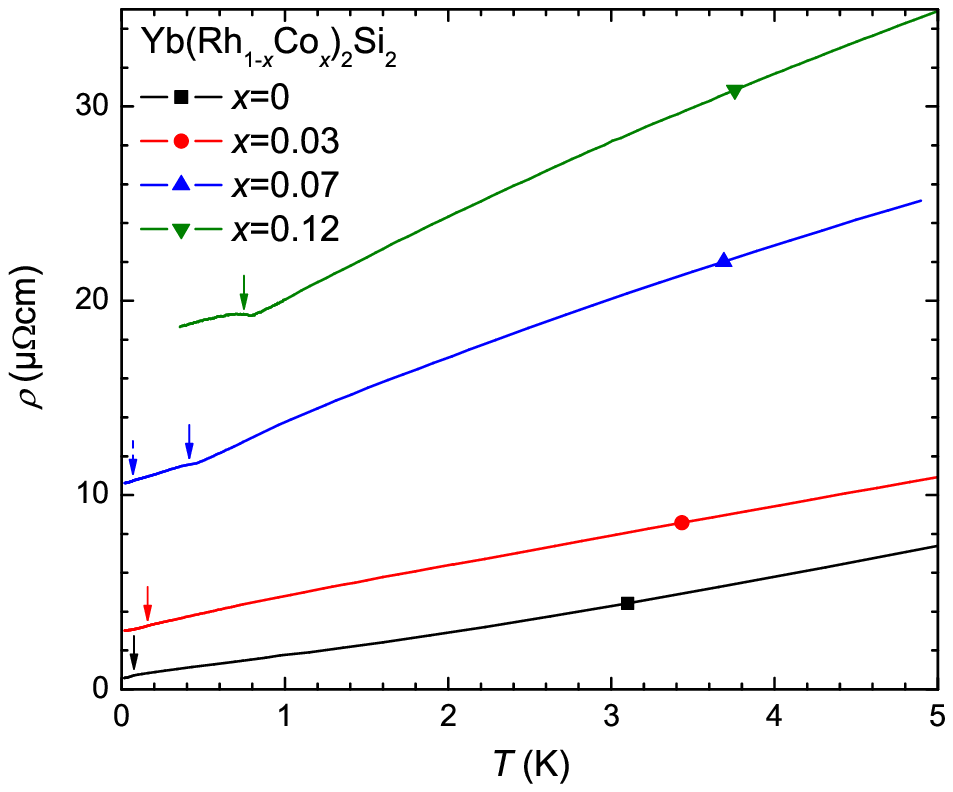}\\
		\includegraphics[width=8.5cm]{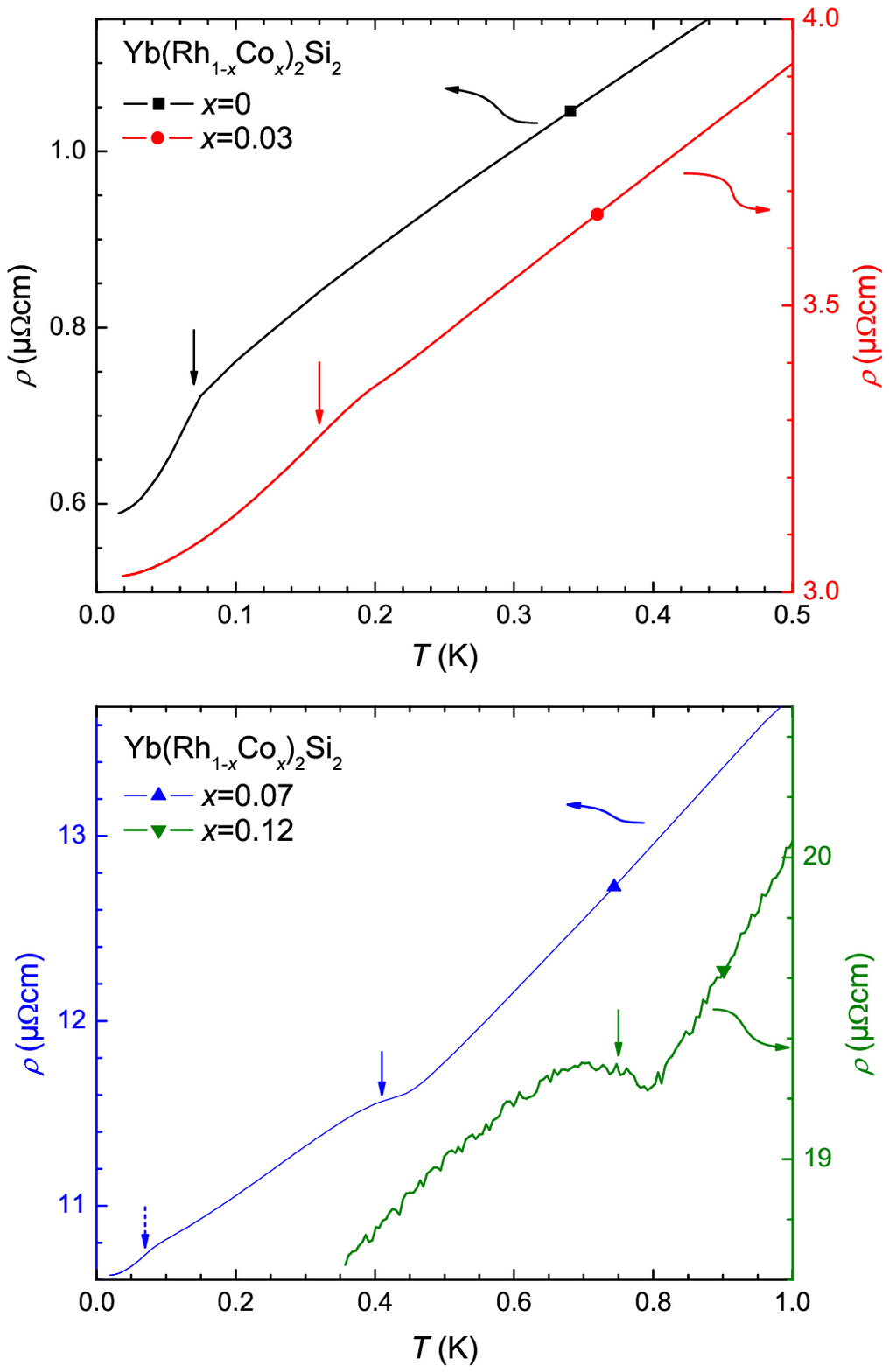}
	\caption{\label{fig:RhovsT_YCxS} (color online) Temperature dependence of the electrical resistivity $\rho(T)$ for various \YCxS\ single crystals. The solid and the dashed arrows indicate the antiferromagnetic ordering temperature \TN\ and the second phase transition temperature \TL, respectively.}
\end{figure}

\begin{figure}
	\begin{center}
		\includegraphics[width=7.6cm]{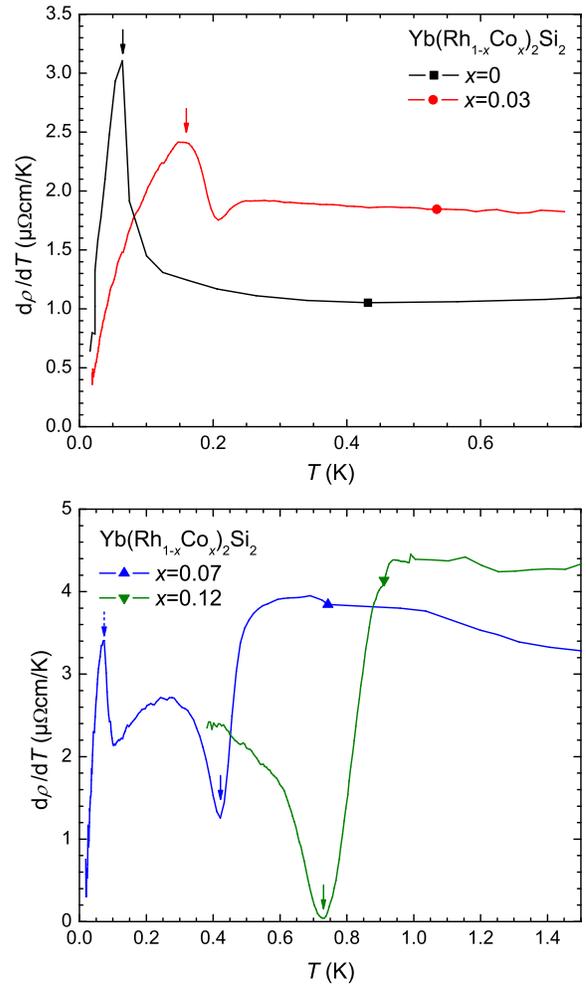}
	\end{center}
	\caption{\label{fig:dRhodTvsT_YCxS} (color online) Derivative of the resistivity with respect to temperature $\dd \rho / \dd T$ for the \YCxS\ single crystals. The solid and the dashed arrows indicate the antiferromagnetic ordering temperature \TN\ and the second phase transition temperature \TL, respectively.}
\end{figure}

\begin{figure}
	\begin{center}
		\includegraphics[width=8.1cm]{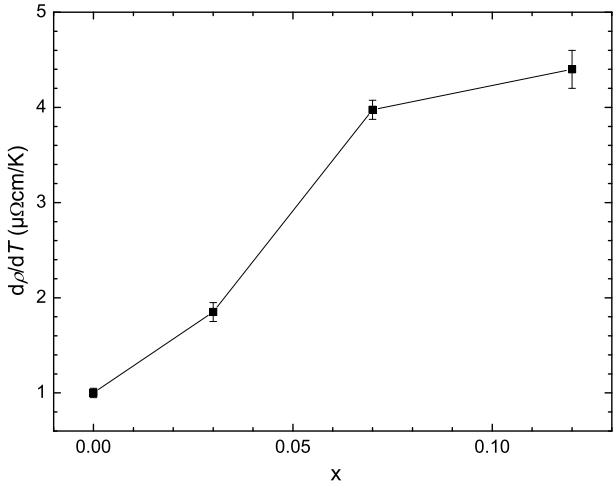}
	\end{center}
	\caption{\label{fig:dRhodTvsX_SCES08} The resistivity slope $\dd \rho / \dd T$ of the linear-in-temperature regime plotted as a function of the Co content $x$.}
\end{figure}

The low-temperature resistivity for \YCxS\ with $x\leq 0.12$ is presented in Fig.~\ref{fig:RhovsT_YCxS}. In accordance to the susceptibility measurements, the antiferromagnetic phase transition is found to be shifted to higher temperatures with increasing Co content. For all samples, a linear temperature dependence is observed above the \Neel\ temperature up to at least \unit{5}\kelvin\ resembling the non-Fermi liquid behavior of the resistivity in \YRS\ \cite{Custers2003}. This behavior is reflected by an almost constant value of the derivative with respect to temperature $\dd \rho / \dd T$ in the corresponding temperature interval as depicted in Fig.~\ref{fig:dRhodTvsT_YCxS}. The slope of $\rho(T)$ is found to increase with increasing Co content as displayed in Fig.~\ref{fig:dRhodTvsX_SCES08}. This might indicate an increase of the fluctuating magnetic moment with increasing Co content. A cross check by a Curie-Weiss analysis of the susceptibility is in agreement to this interpretation although the increasing transition temperature leads to a smaller temperature range which  reduces the reliability of the Curie-Weiss analysis \cite{Westerkamp2009}.
 
\begin{figure}
	\begin{center}
		\includegraphics[width=8.5cm]{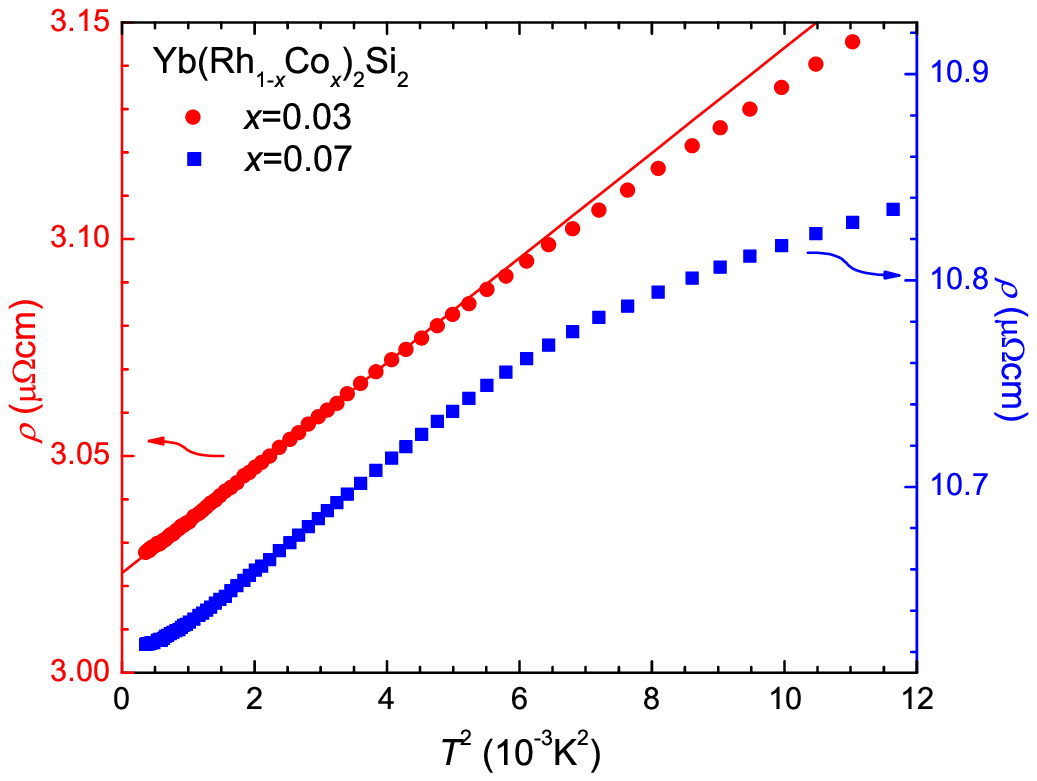}
	\end{center}
	\caption{\label{fig:RhovsT2_YCxS} (color online) Low-temperature resistivity of \YCxS\ with $x=0.03$ (left axis) and $x=0.07$ (right axis) plotted vs. $T^2$. The solid line represents a linear fit to the data of \YCdS .}
\end{figure}

For $x=0.03$ the antiferromagnetic transition is observed as a clear drop. It is best identified in the derivative as a maximum at $\TN=\unit{0.16}\kelvin$. This is at a slightly lower temperature compared to the results of the susceptibility measurements ($\TN=\unit{0.175}\kelvin$) \cite{Westerkamp2008} which might be due to the different criteria used to determine \TN. However, the characteristics of the signature in resistivity resembles those of stoichiometric \YRS\ (shown for comparison in  Fig.~\ref{fig:RhovsT_YCxS}). Below the \Neel\ transition, a quadratic temperature dependence of the resistivity, i.e. $\rho - \rho_0 = A T^2$, was found in a previous study of \YRS\ \cite{Custers2003}. Here, $\rho_0$ is the residual resistivity and $A$ the coefficient of the quadratic term. This behavior is reproduced in \YCxS\ with $x=0.03$ at temperatures below \unit{0.06}\kelvin\ where the data fall on a straight line in the representation vs. $T^2$ as evidenced in Fig.~\ref{fig:RhovsT2_YCxS}. The $A$ coefficient---reflecting the quasiparticle-quasiparticle scattering cross section---is reduced for \YCxS\ with $x=0.03$ compared to \YRS.

In the case of \YCsS, the signature at the AF phase transition is different to \YCxS\ with $x=0.03$ and \YRS. Here, the resistivity exhibits an offset at \TN\ with respect to the linear extrapolation from higher temperatures (\TN\ is marked by a solid arrow in Fig.~\ref{fig:RhovsT_YCxS}). Within the antiferromagnetic phase, an almost linear temperature dependence is found, followed by a second anomaly observed as a drop of the resistivity at \TL\ (cf. dashed arrow in Fig.~\ref{fig:RhovsT_YCxS}). Both features are identified in the derivative $\dd \rho /\dd T$ in Fig.~\ref{fig:dRhodTvsT_YCxS} where a minimum is observed at $\TN=\unit{0.41}\kelvin$ and a maximum is seen at $\TL=\unit{0.07}\kelvin$. The low-temperature signature is assigned to a second magnetic transition previously found in susceptibility measurements \cite{Westerkamp2008}. Although the signature at \TL\ is similar to the one at \TN\ in \YRS\ and \YCxS\ with $x=0.03$, a quadratic temperature dependence is absent for \YCxS\ with $x=0.07$ below \TL\ (cf. Fig.~\ref{fig:RhovsT2_YCxS}). However, it was not possible to find a precise description of $\rho(T)$ in the small temperature range below \unit{45}\mK.

In \YCxS\ with $x=0.12$, an even more pronounced increase is present at the antiferromagnetic transition. From the minimum in the derivative, the transition temperature is determined to be $\TN=\unit{0.75}\kelvin$. A second anomaly was not detected in the investigated temperature range. However, according to susceptibility measurements \cite{Westerkamp2008}, the second transition is expected at $\TL\approx\unit{0.3}\kelvin$, below the limit of our experimental accessible temperature range for this particular set-up.

\begin{figure}
	\begin{center}
		\includegraphics[width=8.5cm]{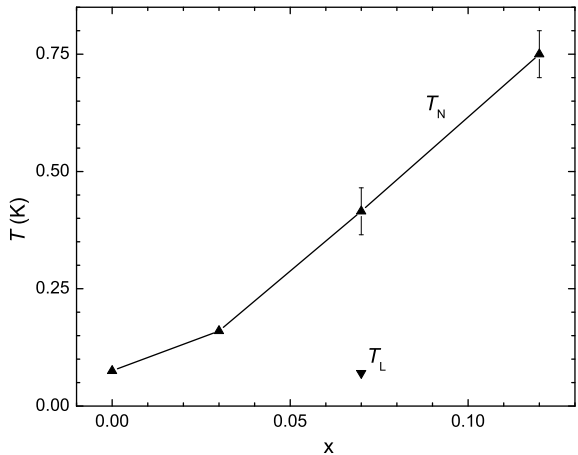}
	\end{center}
	\caption{\label{fig:TvsX_YCxS} The antiferromagnetic transition temperature \TN\ and the second transition temperature \TL\ displayed as a function of Co content $x$. Note that the measurement range for the sample with $x=0.12$ was not extended to below 0.35\,K where the second transition is expected to occur.}
\end{figure}

Summarizing these results, the temperature-chemical composition phase diagram of \YCxS\ is constructed in Fig.~\ref{fig:TvsX_YCxS}. Here, it is seen, that \TN\ increases with increasing Co content. Furthermore, a second transition is observed for the sample with $x=0.07$. Both findings are in good agreement with the results of the susceptibility measurements \cite{Westerkamp2008}. In addition, the increase of the \Neel\ temperature with increasing Co content resembles the evolution of \TN\ under hydrostatic pressure, where the second anomaly is found for pressures $P\geq\unit{1.5}\giga\pascal$ \cite{Mederle2002}. This demonstrates that Co substitution mainly acts as chemical pressure.

Despite the good agreement of the evolution of the transition temperature under hydrostatic pressure compared to that under Co substitution, the actual signature at \TN\ is different for \YCxS\ with Co content $x\geq 0.07$. Whereas, the resistivity under pressure continues to decrease below \TN\ in the whole pressure range \cite{Mederle2002}, the above described offset is observed for \YCxS\ with $x\geq 0.07$. This difference might indicate different magnetic structures for \YCxS\ with $x \geq 0.07$ compared to \YRS\ under an equivalent hydrostatic pressure. The anomaly at \TN\ for $x\geq 0.07$ can be compared with the HF system CeCu$_{5.8}$Au$_{0.2}$. Here, a similar increase at \TN\ was reported for current aligned parallel to one of the components of the magnetic ordering vector $Q$ \cite{Lohneysen1998}. Consequently, the increase could be attributed to the opening of a gap along this direction. On the other hand, for alignment perpendicular to $Q$ a decrease was found in CeCu$_{5.8}$Au$_{0.2}$. Thus, it would be interesting to study the resistivity along different crystallographic directions in \YCxS\ with $x=0.07$.

To conclude, we have shown that the partial substitution of Rh in \YRS\ by smaller Co atoms allows to study the effects of a volume decrease to the ground state properties of \YRS. The induced chemical pressure stabilizes the AF ordering, leading to an increase of the \Neel\ temperature. Furthermore, a second transition is observed for $x=0.07$ in agreement with the results of a previous study of the susceptibility on \YCxS. Pronounced non-Fermi-liquid behavior is detected in a wide temperature range for all concentrations although Co substitution seems to push the system away from the quantum critical point connected with the vanishing magnetic order. Further effort must be undertaken to study the magnetic phases in \YCxS. A detailed investigation of the magnetic field temperature phase diagram of the \YCxS\ series should shed light on the nature of the various energy scales of \YRS.

We would like to acknowledge fruitful discussions with P. Gegenwart and T. Westerkamp. This work was partially supported by the DFG research group 960 ``Quantum Phase Transitions''.


\begin{thebibliography}{99}
\bibitem{Trovarelli2000b} O. Trovarelli et al., Phys. Rev. Lett. {\bf 85} (2000) 626.
\bibitem{Custers2003} J. Custers et al., Nature {\bf 424} (2003) 524.
\bibitem{Kuchler2003} R. K{\"u}chler et al., Phys. Rev. Lett. {\bf 91} (2003) 066405.
\bibitem{Millis1993} A. J. Millis, Phys. Rev. B {\bf 48} (1993) 7183.
\bibitem{Moriya1995} T. Moriya and T. Takimoto, J. Phys. Soc. Jpn. {\bf 64} (1995) 960.
\bibitem{Paschen2004} S. Paschen et al., Nature {\bf 432} (2004) 881.
\bibitem{Si2001} Q. Si et al., Nature {\bf 413} (2001) 804.
\bibitem{Gegenwart2007} P. Gegenwart et al., Science {\bf 315} (2007) 969.
\bibitem{Mederle2002} S. Mederle et al., J. Phys.: Condens. Matter {\bf 14} (2002) 10731.
\bibitem{Westerkamp2008} T. Westerkamp et al., Physica B {\bf 403} (2008) 1236.
\bibitem{Westerkamp2009} T. Westerkamp private communication.
\bibitem{Gegenwart2008} P. Gegenwart et al., Nat. Phys. {\bf 4} (2008) 186.
\bibitem{Lohneysen1998} H. v. L{\"o}hneysen et al., Eur. Phys. J. B {\bf 5} (1998) 447.

\end{thebibliography}
\end{document}